# Room temperature ferromagnetism of monolayer chromium telluride with perpendicular magnetic anisotropy

Rebekah Chua[1,2], Jun Zhou[2], Xiaojiang Yu[3], Wei Yu[4], Jian Gou[2], Rui Zhu[2], Lei Zhang[2], Mark B. H. Breese[2,3], Wei Chen,[2,4,6] Kian Ping Loh[1,4], Yuan Ping Feng[1,2], Ming Yang[5*], Yu Li Huang[2,6*], Andrew T. S. Wee[1,2*]

[1]*NUS Graduate School for Integrative Sciences & Engineering (NGS), University Hall, Tan Chin Tuan Wing, 21 Lower Kent Ridge, Singapore 119077, Singapore*
[2]*Department of Physics, National University of Singapore, 2 Science Drive 3, Singapore 117551, Singapore*
[3]*Singapore Synchrotron Light Source (SSLS), National University of Singapore, 5 Research Link, Singapore 117603, Singapore*
[4]*Department of Chemistry, National University of Singapore, 2 Science Drive 3, Singapore 117542, Singapore*
[5] *Department of Applied Physics, The Hong Kong Polytechnic University, Hung Hom, Hong Kong, China*
[6] *Joint School of National University of Singapore and Tianjin University, International Campus of Tianjin University, Fuzhou 350207, China*

*\*E-mail: mingyang@polyu.edu.hk; hyl@tjufz.org.cn; phyweets@nus.edu.sg*

**Abstract**

The realization of long-range magnetic ordering in two-dimensional (2D) systems can potentially revolutionize next-generation information technology. Here, we report the successful fabrication of crystalline $Cr_3Te_4$ monolayers with room temperature ferromagnetism. Using molecular beam epitaxy, the growth of 2D $Cr_3Te_4$ films with monolayer thickness is demonstrated at low substrate temperatures (~100°C), compatible with Si CMOS technology. X-ray magnetic circular dichroism measurements reveal a Curie temperature ($T_c$) of ~344 K for the $Cr_3Te_4$ monolayer with an out-of-plane magnetic easy axis, which decreases to ~240 K for the thicker film (~ 7 nm) with an in-plane easy axis. The enhancement of ferromagnetic coupling and the magnetic anisotropy transition is ascribed to interfacial effects, in particular the orbital overlap at the monolayer $Cr_3Te_4$/graphite interface, supported by density-functional theory calculations. This work sheds light on the low-temperature scalable growth of 2D nonlayered materials with room temperature ferromagnetism for new magnetic and spintronic devices.

**Introduction**

The family of two-dimensional (2D) crystals has been expanding rapidly over the past decade, exhibiting a wide variety of novel electronic and optical properties, and more recently magnetism.[1-3] This is because long-range magnetism is not predicted to manifest in 2D systems due to the enhanced thermal fluctuations, as predicted by the Mermin-Wagner theorem.[4] In 2017, breakthroughs in 2D magnetism were achieved in atomically thin layers of $CrI_3$[5] and $Cr_2Ge_2Te_6$,[6] showing that magnetic anisotropy can stabilize the long-range magnetic ordering with an excitation gap opening. Subsequently, other 2D materials with exotic magnetic properties, e.g., $Fe_3GeTe_2$,[7] $FePS_3$,[8] $VSe_2$,[9-13] and $MnSe_2$[14], have been reported. However, challenges remain in identifying true monolayer 2D magnets with room temperature ferromagnetic ordering, and large-area low temperature controllable growth. Their novel properties and emergent phenomena in the 2D limit require validation for applications in next generation memory and information technology, spintronics, and quantum computing.

Chromium chalcogenides ($Cr_mX_n$, X = S, Se and Te), an emergent class of non-layered materials with robust magnetic ordering, have attracted increasing research interest. For example, epitaxial $Cr_2S_3$,[15] $CrSe$,[16] $CrTe_2$,[17] $CrTe$,[18] and $Cr_2Te_3$[19-21] thin films with thicknesses down to several nanometers have been grown by chemical vapour deposition (CVD) and/or molecular beam epitaxy (MBE), showing high Curie temperature $T_C$ (e.g., ~305 K for 8.7 nm $CrTe_2$ thin film[17]), and promising electrical transport properties (e.g., unconventional anomalous Hall effect[18]) for spintronic devices. The successful synthesis of 2D $Cr_2Te_3$[22] and $CrTe_2$[23] with monolayer thickness have also been recently reported, but characterization of their magnetic properties is lacking.

Herein, we report the room temperature ferromagnetism of 2D $Cr_3Te_4$ in the monolayer form. Using the MBE method, we successfully grew crystalline $Cr_3Te_4$ monolayers on graphite at low substrate temperatures, i.e., ~100°C. The growth mode transits from 2D to 3D at elevated substrate temperatures above 300°C. X-ray magnetic circular dichroism (XMCD) measurements reveal that the magnetic properties of 2D $Cr_3Te_4$ are modulated by the film thickness. Thicker $Cr_3Te_4$ films (~7 nm) have a $T_C$ lower than room temperature (RT) along its in-plane easy axis (~240 K), while monolayer $Cr_3Te_4$ displays RT ferromagnetism with $T_C$ up to ~344 K, and perpendicular magnetic anisotropy. Density-functional theory (DFT) calculations suggest that the transition of the magnetic easy axis, and the enhancement of ferromagnetic coupling in the $Cr_3Te_4$ monolayer is ascribed to interfacial effects at the monolayer $Cr_3Te_4$/graphite interface. The low temperature growth process, and the realization of room-temperature magnetism in 2D $Cr_3Te_4$ will facilitate the practical applications of 2D magnets in spintronic devices and high-density memory devices, compatible with Si CMOS technology.

**Results and Discussion**

Bulk chromium tellurides ($Cr_mTe_n$) are a group of materials with similar structures, which can be taken as derivates of the parent CrTe. CrTe has a nonlayered structure consisting of octahedral $CrTe_2$ layers (an analogy to 1T $MoS_2$ monolayer) connected by intercalated Cr atoms bonding with neighbouring Te atoms in trigonal prism. With different amounts of Cr vacancies in the intercalated Cr layers, a range of $Cr_mTe_n$ with varied stoichiometric ratios such as CrTe, $Cr_2Te_3$, $Cr_3Te_4$, and $Cr_5Te_8$, can be formed.[24,25] In particular, bulk $Cr_3Te_4$ as shown in Figure 1a and b has a Te-Cr-Te-Cr quadruple non-layered structure, with 50% Cr vacancies in the intercalated layer (space group P2/m). Bulk $Cr_3Te_4$ is ferromagnetic with an averaged magnetic moment predicted to be 3.32 $\mu_B$ per Cr atom by our DFT calculations, which is close to the previously reported values.[26] It is noted that bulk $Cr_3Te_4$ is not subject to a T'-like phase transition as indicated by the negligible distance difference between neighbouring Cr-Cr atoms (Table 1). When the thickness decreases to the monolayer limit (5 atom layers, i.e., Te-Cr-Te-Cr-Te), the composition ratio changes to Cr:Te =1:2 (Fig. 1c-d). Remarkably, monolayer $Cr_3Te_4$ shows a T'-like phase transition, leading to periodic lattice distortion, i.e., a 0.32 Å difference of between neighbouring Cr-Cr bonds (Fig. 1d). The averaged magnetic moment of Cr in ML $CrTe_2$ is 3.15 $\mu_B$ per atom, close to that of bulk. Interestingly, the magnetic anisotropy changes from an in-plane easy magnetic axis of around 0.50 meV/f.u. in bulk $Cr_3Te_4$ to an out-of-plane easy axis of around -0.40 meV/f.u in freestanding monolayer $Cr_3Te_4$, the latter of which can be further enhanced to -2.17 meV/f.u. by the graphene substrate. The unexpected switching of the easy axis of magnetization between the bulk and monolayer $Cr_3Te_4$ could be a result of the thickness effect. In the monolayer limit, $Cr_3Te_4$ has different local coordination environments and a stronger substrate effect, and hence different interatomic exchange, resulting in preferential perpendicular magnetic anisotropy.

To realize the 2D growth of the non-layered $Cr_3Te_4$ crystal, MBE was selected as precise growth at the monolayer level is achievable by controlling source fluxes and substrate temperature (see Methods). The morphology of the MBE-grown $Cr_3Te_4$ ultrathin films on graphite varies significantly with the substrate temperature by keeping the evaporation rate constant. Figure 2a-d show the large-scale topographic images (500 × 500 $nm^2$) of $Cr_3Te_4$ thin films (before the substrates were fully covered) grown at different substrate temperatures, 400°C, 300°C, 200°C and 100°C, respectively. The as-grown $Cr_3Te_4$ flakes are typically regular hexagonal shape at high temperature (e.g., 400 °C and 300 °C), and irregular dendritic shape at low temperature (e.g., 100°C). The thickness of the $Cr_3Te_4$ flakes decreases gradually with the substrate temperature, as revealed by the lateral profiles in Fig. 2e-h. Typically, the thickness of the $Cr_3Te_4$ flakes is ~4.5 nm at 400°C (Fig. 2e), 2.1 – 3.0 nm at 300°C (Fig. 2f), 0.9 – 2.1 nm at 200°C (Fig. 2g), and 0.9 – 1.5 nm at 100°C (Fig. 2h). For the samples grown at 100°C and 200°C, the height of the first layer is only 0.93±0.2 nm, indicating the successful growth of 2D $Cr_3Te_4$ with monolayer thickness. This height consists of a $Cr_3Te_4$ monolayer (a Te-Cr-Te-Cr-Te quintuple layer) in addition with an interfacial separation between the $Cr_3Te_4$ film and the graphite substrate. The subsequent layers increase by an increment of ~0.6 nm or ~0.3 nm, which corresponds to a Te-Cr-Te-Cr quadruple layer (Fig. 1) and a Te-Cr bi-atomic layer, respectively. The less observed ~0.3 nm increment is unique for such non-layered materials.

Clearly, the $Cr_3Te_4$ films tend to grow in 3D island (Volmer-Weber) mode at higher substrate temperatures (300°C - 400°C), while in layer-plus-island (Stranski-Krastanov) mode at lower temperatures (100°C - 200°C) (see Figure S1 in Supplementary Information, SI). Since the $Cr_3Te_4$ crystal is nonlayered and possesses isotropic covalent bonding in all three dimensions, preferential growth in 3D island mode occurs under thermodynamic equilibrium conditions, e.g., at higher substrate temperatures. The interfacial interactions between the $Cr_3Te_4$ film and graphite substrate could play a critical role in promoting anisotropic growth, facilitating 2D growth mode at lower temperatures.

Further studies were carried out to investigate 2D $Cr_3Te_4$ films grown at 100°C under optimized conditions. Figure 3a and 3b show the topological images of $Cr_3Te_4$ monolayer and thicker film (~ 7 nm), respectively. Fig. 3c represents a typical STM image with atomic resolution recorded at monolayer region. A quasi-hexagonal lattice structure with unidirectional 1D modulation is observed, similar to that commonly observed in 1T'-TMDs.[27,28] The unit cell is highlighted by a red rectangle, with *a* = 6.9 ± 0.1 Å and *b* = 4.0 ± 0.1 Å, agreeing well with the calculated lattice constant of $Cr_3Te_4$ (Table 1). The 1D modulation can be easily captured in the corresponding Fast Fourier Transform (FFT) image given in Fig. 3e, attributing to the peaks highlighted by the white circles. This 1D superstructural modulation is still visible in the non-contact atomic force microscopy (nc-AFM) image shown in Fig. 3d, indicating that it is probably due to lattice distortion. Variable temperature STM measurements reveal that the 1D modulation remains up to 206 K (Figure

S2 in SI), indicating that it might not arise from a charge-density wave. The same atomic structure is observed in bi- and multi-layer films, confirming the uniformity of the MBE-grown $Cr_3Te_4$ samples. Fig. 3f shows the simulated STM image of monolayer $Cr_3Te_4$ on a single-layer graphene substrate (approximation of bulk graphite), based on the model presented in Fig. 1c-d, consistent with the experimental STM image in Fig. 3c. The 1D modulation, observed both experimental and theoretically, is ascribed to the surface fluctuation as well as the bond distortion arising from periodic Cr vacancies in the $Cr_3Te_4$ monolayer (Figure S3 in SI).

Photoemission spectroscopy (PES) measurements based on synchrotron radiation were performed to determine the chemical composition and the electronic structure of the as-grown samples. Fig. 3g-h show the Cr 2p and Te 3d core-level spectra of $Cr_3Te_4$ thin (~1 ML) and thicker (~ 7 nm) films, respectively. The two broad peaks located at ~572 and ~582 eV are assigned to Te-$3d_{5/2}$+Cr-$2p_{3/2}$ and Te-$3d_{3/2}$+Cr-$2p_{1/2}$, respectively,[29,30] since the binding energies of the Cr 2p and Te 3d core levels are close in energy. We fit the spectra with three component peaks, i.e., Te-$3d_{5/2}$ (Te-$3d_{3/2}$), Cr-$2p_{3/2}$ (Cr-$2p_{1/2}$) and Te from $Te_xO_y$ for the peak located at lower (higher) binding energy. The binding energy and full width at half maximum (FWHM) for each component are given in Table S1 and Table S2 in SI. The stoichiometric ratio between Cr and Te elements is estimated to be 0.56 ± 0.02 in Fig. 3g for the $Cr_3Te_4$ monolayer sample, consistent with the proposed mode in Fig. 1d with Cr:Te = 1:2. In Fig. 2h, the estimated Cr:Te ratio increases to 0.70 ± 0.03 for the thicker film, which is close to the stoichiometry of the bulk $Cr_3Te_4$. The small $Te_xO_y$ components are due to the residual of the capping Te layers, used to encapsulate the $Cr_3Te_4$ samples during the *ex-situ* transfer, and removed by thermally annealing at 250°C - 300°C before PES measurements (Figure S4 in SI).

Fig. 3i shows a d*I*/d*V* spectrum acquired on $Cr_3Te_4$ monolayer showing a "V" shape feature around the Fermi level, indicating that it is a gapless material. Similar feature is also observable on $Cr_3Te_4$ films with various thickness. The metallic nature is confirmed by PES measurements at the valence band region, where sharp Fermi edges are observed for both monolayer and thicker films (Figure S5a in SI). Furthermore, from the secondary cutoff, the work function (WF) of the monolayer $Cr_3Te_4$ is determined to be 5.19 eV, which slightly increases to 5.49 eV for the thicker film (Figure S5b in SI). Since the WF of a clean graphite is only 4.32 eV (Figure S6 in SI), the total increase of 0.87 eV (1.17 eV) with a $Cr_3Te_4$ monolayer (thicker) film indicates a possible charge transfer from the graphite substrate to n-dope the $Cr_3Te_4$ film.

To examine the magnetic properties of the $Cr_3Te_4$ films, elemental-specific X-ray adsorption spectroscopy (XAS) and XMCD measurements were performed at the Cr $L_{2,3}$ edge at 78 K – 350 K by the total photoelectron yield method. Figure 4a-d show the XAS spectra measured under a magnetic field of $\mu H = \pm 1$ T, and the XMCD spectra corresponding to the difference of the two XAS pairs ($\mu^+ - \mu^-$) for both normal incidence (NI) and grazing incidence (GI). Here, $\mu^+$ and $\mu^-$ represent the absorption spectra for the spin direction of Cr 3d majority electrons parallel and antiparallel to the photon helicity. The absorption from Cr-$2p_{3/2}$ and Cr-$2p_{1/2}$ core levels present as multiplet structures around the photon energies of 576 eV and 586 eV, respectively. The observed structures are comparable with those previously reported for $Cr^{2+}/Cr^{3+}$ compounds, serving as a spectroscopic fingerprint of $Cr_3Te_4$.[31,32] For the $Cr_3Te_4$ monolayer, the XMCD curves in Fig. 4a-b demonstrate a non-zero signal up to 330 K for both NI and GI. The signal is most prominent at 78 K, and gradually decreases with increasing temperature, and finally vanishes at 350 K. The XMCD signal at NI is stronger than that at GI, indicating magnetic anisotropy in the monolayer $Cr_3Te_4$. In comparison, the XAS and XMCD spectra for the $Cr_3Te_4$ thicker film (~7 nm) are shown in Fig. 4c-d. The XMCD signals for both NI and GI are comparable at low temperatures, indicating that the magnetic anisotropy is less pronounced in the thicker film. The XMCD signal diminishes as the temperature increases, up to around 250 K at NI, and 300 K GI.

Further analyses of the XMCD spectra based on sum rules analysis[33] were carried out to estimate the magnetic moments of the Cr element. Fig. 4e-f show the summary of the magnetism of the $Cr_3Te_4$ monolayer and thicker films as a function of temperature. In the thicker film (Fig. 4e), the total magnetic moment ($m_t = m_s + m_L$) in both NI and GI directions are similar from 78 K to 300 K. $m_t$ is of ~1.23 $\mu_B$ per atom in the thicker $Cr_3Te_4$ film at 78 K, and almost vanishes at 250 K. Fitting the experimental data by $M(T) \propto (1 - \frac{T}{T_C})^\gamma$, we yield $T_c$ of 248 ± 5 K in the GI direction, and 235 ± 5 K in the NI direction, both slightly below room temperature. The thicker film demonstrates less magnetic anisotropy, and its easy axis is parallel to the sample surface. On the other hand, the $Cr_3Te_4$ monolayer shows pronounced perpendicular magnetic anisotropy, where the total magnetic moment at NI is much larger than that at GI. At 78 K, the estimated value is ~0.79 $\mu_B$ per Cr atom at NI, but only ~0.16 $\mu_B$ per atom at GI; at 300 K, the value decreases to ~0.09 $\mu_B$ at NI and ~0.05 $\mu_B$ at GI. Hence, the easy axis for spontaneous magnetization is normal to the sample surface of the monolayer $Cr_3Te_4$. From the fitting M-T curves in Fig. 4f, $T_c$ is around 344 ± 6 K for the $Cr_3Te_4$ monolayer. The existence of ferromagnetism above room temperature in the $Cr_3Te_4$ monolayer are confirmed by magnetic force microscopy (MFM) (Figure S7 in SI). The observed enhancement of ferromagnetic coupling and the magnetic anisotropy transition in monolayer $Cr_3Te_4$ is consistent with our DFT prediction (Table 1).

Further DFT calculations were performed to understand the underlying physics of the thickness-dependent magnetic properties. Figure 5a shows that the spin density is dominantly around the Cr atoms without contribution from graphene substrate. Although the interface spacing between $Cr_3Te_4$ and graphene is as large as 3.5 Å within the range of van der Waals bonding, the interaction between them is non-trivial. Significant charge redistribution is observed in Fig. 5b. The charge density is depleted on the proximity of graphene, and the excess charge density is accumulated on the bottom of $Cr_3Te_4$ monolayer, indicating that the charge transfers from graphene into $Cr_3Te_4$. Further Bader analysis suggests about 0.1 e$^-$ transfers from graphene into the monolayer $Cr_3Te_4$. This is ascribed to interfacial electrostatic interaction. Fig. 5c shows the in-plane averaged electrostatic potential across the $Cr_3Te_4$/graphene interface, from which we can obtain the WF of $Cr_3Te_4$ and graphene, respectively. The calculated WF is 4.3 eV for graphene, and 5.3 eV for $Cr_3Te_4$ monolayer, consistent with our experimentally measured WF data. The WF difference between them leads to the charge transfer when forming $Cr_3Te_4$/graphene heterointerface.

From the PDOS shown in Fig. 5d, we can also see that the Dirac cone of graphene is shifted to higher energy, indicating a hole-doping in graphene due to charge transfer to $Cr_3Te_4$. Besides, the PDOS of Cr atoms near the interface and C atoms show a similar profile around the Fermi level, which clearly suggests orbital overlap between them, especially between $d_{xz}$, $d_{yz}$, and $d_{z2}$ orbitals of Cr and $p_z$ orbital of graphene as mediated by the Te $p_z$ orbital in between. The $p_z$ orbital of graphene may pin the direction of the magnetic moments on Cr atoms, leading to an enhanced out-of-plane magnetic easy axis.[34] We note that the Cr atoms show two different magnetic moments of 3.20 and 3.10 $\mu_B$, respectively, indicating the different oxidization states of the Cr atoms. The double exchange mechanism can be applied in such a situation. The parallel alignment (ferromagnetic coupling) of the magnetic moments on Cr ions allows the hopping of electrons between the Cr ions with different oxidation state, which reduces the kinetic energy.[35] Based on our experiments and DFT calculations, we suggest that the 2D growth mode at low substrate temperature (100°C), the transition of out-of-plane magnetization easy axis, and the high Curie temperature for the $Cr_3Te_4$ monolayer on graphite, are due to the interfacial interaction induced orbital overlap.

We further perform Monte Carlo (MC) simulations with the Metropolis algorithm[36,37] to estimate the Curie temperature of the $Cr_3Te_4$ monolayer. Since the monolayer $Cr_3Te_4$ has an out-of-plane easy magnetic axis and the magnetic anisotropy is large (2.17 meV/f.u.), we can apply the Ising model for Monte Carlo simulations (more details of the method are given in SI). As shown in Fig. 4g, both the temperature-dependent magnetization and magnetic susceptibility suggest a $T_C$ of 300 K for $Cr_3Te_4$ monolayer on graphene, in good agreement with our experimental results. A validation test with the same MC simulation was also performed on $CrI_3$ monolayer, in which the match between the simulated $T_C$ of 52 K and the experimentally reported value of 45 K[5] indicates the reliability of our MC simulations.

In conclusion, 2D ferromagnetic $Cr_3Te_4$ with room temperature ferromagnetism and strong perpendicular magnetic anisotropy have been demonstrated. The $Cr_3Te_4$ films with monolayer thickness are successfully fabricated using MBE growth methods at low substrate temperatures (100°C). Magnetic measurements demonstrate that monolayer $Cr_3Te_4$ has a $T_c$ of ~344 K (above room temperature) and an out-of-plane easy axis, while the thicker film (~ 7 nm) has a $T_c$ of ~240 K with an in-plane easy axis. DFT calculations support our experimental observations and reveal that the non-trivial interactions between monolayer $Cr_3Te_4$ and the graphite substrate results in an interfacial interaction, which induces a perpendicular orbital overlap that stabilizes the out-of-plane magnetic ordering. The realisation of 2D magnetic materials with room temperature ferromagnetism is important for next-generation novel magnetic devices. Furthermore, low temperature 2D MBE growth (100°C) is compatible with Si CMOS technology, opening up new opportunities for integrating 2D spintronic devices into Si based electronic architectures.

## Methods

### $Cr_3Te_4$ growth

The $Cr_3Te_4$ films were grown on commercial highly orientated pyrolytic graphite (HOPG) in an MBE chamber attached to the Omicron LT-STM system with a base pressure of $1 \times 10^{-9}$ mbar. The HOPG substrates were freshly exfoliated and then annealed at 500°C for more than 1 hour before MBE growth. Elemental Cr and Te were used as effusion source materials. During the deposition, the Te:Cr ratio was kept high (~10:1) to maintain Te overpressure since it is more volatile. The graphite substrate was held at elevated temperature (e.g., 100°C to 400°C) to provide sufficient adatom diffusion energy for surface atom mobility to reach near thermodynamic equilibrium. LT-STM measurements were done *in-situ*. For *ex-situ* XPS/XAS measurements, Te capping layers were deposited onto the $Cr_3Te_4$ surface to prevent contamination in ambient environment during the transport. The Te capping layers were removed by annealing at 250°C for 30 minutes.

**LT-STM and nc-AFM measurements**

LT-STM and nc-AFM studies were performed at 77 K in an Omicron UHV system interfaced to a Nanonis controller equipped with STM/qPlus sensor. Electrochemically etched tungsten tips were used, where a bias voltage was applied on it, and the sample holder was grounded. The STM images were acquired under a constant current mode. For nc-AFM imaging, the constant-height mode with an oscillation amplitude of 10 nm was used to record the frequency shift ($\Delta f$) of the qPlus resonator (sensor frequency $f_0 \approx 24$ kHz, $Q \approx 8000$). A lock-in technique was used to measure the d$I$/d$V$ spectra, with a modulation of 625 Hz and 30 mV.

**PES and XAS measurements**

PES, XAS and XMCD measurements were carried out at the SINS beam line of the Singapore Synchrotron Light Source (SSLS). The incident photon energy was 700 eV and the energy resolution was 0.2 eV for the core level spectra. The XAS and XMCD at V $L_{2,3}$ edge at room temperature were obtained by recording the sample current (total yield mode) as a function of photon energy. Elliptically polarized light with a degree of circular polarization (DCP) = 80 and an energy resolution of 0.3 eV was employed for the XMCD measurements. To measure the in-plane and out-of-plane spin and orbital magnetic moments, the light was incident at a grazing or normal angle, respectively, from the sample surface, with its propagation direction along the sample in-plane or out-of-plane magnetization direction. An external magnetic field of ±1 T was applied to magnetize the sample along the in-plane or out-of-plane direction. The XMCD was done by changing the direction of applied magnetic field while keeping the helicity of the light.

**DFT and Monte Carlo Simulations**

The first-principles calculations were performed based on the projector augmented-wave method[38] as implemented in the Vienna ab initio simulation package (VASP).[39] The exchange-correlation interaction was treated with the Perdew–Burke–Ernzerhof form of the generalized gradient approximation (GGA).[40] The Hubbard U of 3.2 eV was adopted to deal with the strongly correlated effect between Cr 3$d$ electrons.[41] The plane-wave cutoff energy was set to 500 eV. The lattice parameters and atomic positions were fully relaxed until the energy and force on each atom were converged to less than $1 \times 10^{-6}$ eV and $1 \times 10^{-3}$ eV/Å, respectively. The interface of $Cr_3Te_4$ monolayer/graphene was modelled by placing ($\sqrt{3} \times 1$) monolayer $Cr_3Te_4$ supercell on ($3 \times \sqrt{3}$) graphene supercell, in which about 3.7% tensile strain was applied on $Cr_3Te_4$ monolayer. In this model, we used graphene monolayer to represent graphite substrate, which does not affect our main results due to the weak interlayer interaction between graphite layers. For the interface calculations, Van der Waals correction was considered using the method of Grimme (DFT-D3).[42] The STM images were simulated using the Tersoff-Hamann method [43]. The Bader charge analysis was used to estimate the number of valence charges.[44] For the comparison of magnetic anisotropy energy, the spin-orbital coupling effect was included in the calculations with a higher accuracy of $1 \times 10^{-8}$ eV. In all calculations, the dipole correction was applied. For the comparison of magnetic anisotropy energy, the spin-orbital coupling effect was included in the calculations with a higher accuracy of $1 \times 10^{-8}$ eV. In all calculations, the dipole correction was applied.

The Monte Carlo simulations were performed based on Ising spin Hamiltonians using 60×60×1 supercells with periodic boundary conditions. For each temperature, $5 \times 10^7$ steps were used to reach thermal equilibrium. The first $2.5 \times 10^7$ steps were discarded, and the last $2.5 \times 10^7$ steps were used to calculate the temperature dependent physical quantities.

## Acknowledgements

This work was funded by the NRF of Singapore (grant no. R-144-000-405-281) and MOE Tier 2 grant MOE2016-T2-2-110. Y.L.H. acknowledges the support from the National Natural Science Foundation of China (project No.: 12004278). M.Y. acknowledges the start-up funding support (project ID: 1-BE47) from The Hong Kong Polytechnic University. We acknowledge Centre for Advanced 2D Materials and Graphene Research at National University of Singapore, and National Supercomputing Centre of Singapore for providing computing resource.

**Table 1** DFT calculated structural and magnetic properties for bulk $Cr_3Te_4$, freestanding monolayer $Cr_3Te_4$ and $Cr_3Te_4$/graphene heterostructure, respectively. $m_{Cr}$ and $m_{Te}$ are the averaged magnetic moments on Cr and Te, respectively. $a$ and $b$ are the in-plane lattice parameters, while $d$ is the thickness of slab in u.c.. $d_{bond}$ is the distance difference between neighbouring Cr-Cr atoms. $E_{mae}$ is the magnetic anisotropy energy, which is defined by the energy difference between the magnetic moments on Cr atoms aligning along [001] and [100] directions.

|  | $m_{Cr}$ ($\mu_B$) | $m_{Te}$ ($\mu_B$) | $a$ (Å) | $b$ (Å) | $d$ (Å) | $d_{bond}$ (Å) | $E_{mae}$ (meV/f.u.) | Magnetic order |
|---|---|---|---|---|---|---|---|---|
| $Cr_3Te_4$-bulk | 3.32 | -0.26 | 6.97 | 4.06 | 6.42 | -0.02 | 0.50 | FM$^{\parallel}$ |
| $Cr_3Te_4$-ML | 3.15 | -0.21 | 6.86 | 3.96 | 6.33 | -0.32 | -0.40 | FM$^{\perp}$ |
| $Cr_3Te_4$-ML/Graphene | 3.14 | -0.23 | -- | -- | -- | -0.33 | -2.17 | FM$^{\perp}$ |

$^{\parallel}$In-plane anisotropy. $^{\perp}$Out-of-plane anisotropy. FM, ferromagnet.

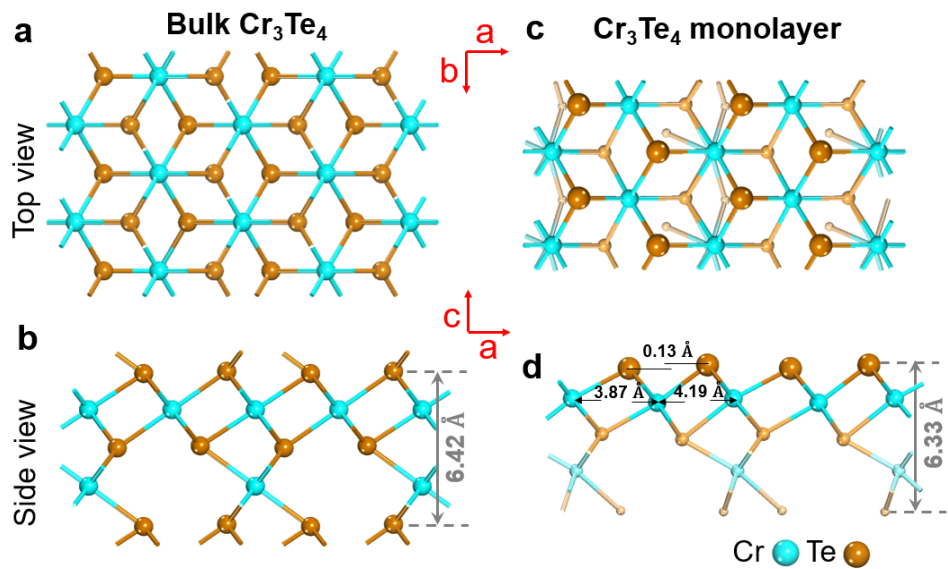

**Figure 1** Atomic models of bulk $Cr_3Te_4$ crystal and monolayer $Cr_3Te_4$ with Te termination (2D $Cr_3Te_4$). (a) Top view and (b) side view of bulk $Cr_3Te_4$, and (c) top view and (d) side view of 2D $Cr_3Te_4$. The different size and color of balls indicate the atoms in different atomic layers.

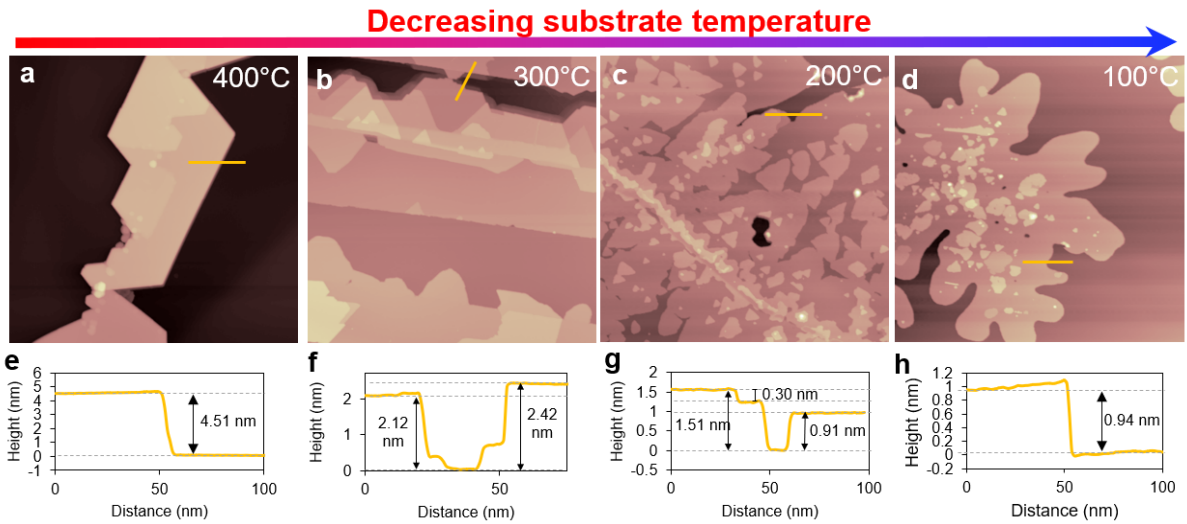

**Figure 2** Cr$_3$Te$_4$ thin films grown at various substrate temperature. (a-d) Large-scale STM images (500 × 500 nm$^2$) reveal the temperature-dependent morphology of Cr$_3$Te$_4$ films grown at: (a) 400°C ($V_{Tip}$ = 1.0 V, $I_{Tip}$ = 20 pA), (b) 300°C ($V_{Tip}$ = 1.0 V, $I_{Tip}$ = 20 pA), (c) 200°C ($V_{Tip}$ = 1.5 V, $I_{Tip}$ = 20 pA), and (d) 100°C ($V_{Tip}$ = 2.0 V, $I_{Tip}$ = 10 pA). (e-h) Line profiles correspond to the yellow line in panels (a-d), respectively, revealing the different thicknesses of the Cr$_3$Te$_4$ flakes. The averaged thicknesses of the Cr$_3$Te$_4$ films are 4.45 nm, 2.35 nm, 1.43 nm and 1.0 nm, for (a-d).

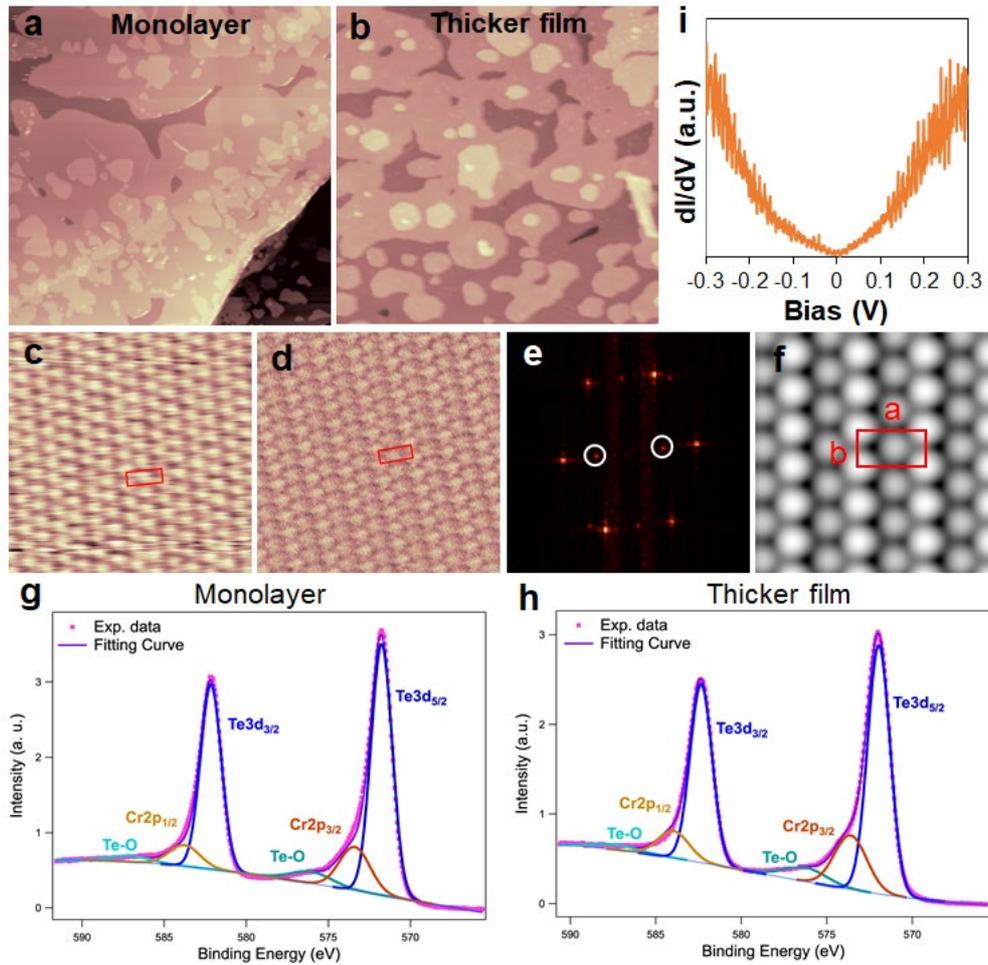

**Figure 3** Crystal structures and electronic properties of Cr$_3$Te$_4$ monolayer and thick film. (a-b) Large-scale STM image of (a) Cr$_3$Te$_4$ monolayer (1 u.c.) and (b) thick film (10 u.c.) (a, 300 × 300 nm$^2$; $V_{Tip}$ = 2.0 V, $I_{Tip}$ = 10 pA; b, 150 × 150 nm$^2$; $V_{Tip}$ = 2.0 V, $I_{Tip}$ = 10 pA). (c) A typical STS spectroscopy of Cr$_3$Te$_4$ films (setting points: $V_{Tip}$ = 0.15 V, $I_{Tip}$ = 200 pA). (d) Atomically resolved STM (5 × 5 nm$^2$; $V_{Tip}$ = 0.05 V, $I_{Tip}$ = 588 pA), and (e) a constant-height nc-AFM image (5 × 5 nm$^2$) reveal 1D unidirectional modulation in monolayer Cr$_3$Te$_4$. (f) FFT image obtained from the atomically resolved STM image in (d). (g) Simulated STM image of monolayer Cr$_3$Te$_4$ supported by a single-layer graphene. (h-i) High-resolution Cr 2p and Te 3d core level spectra of (h) Cr$_3$Te$_4$ monolayer and (i) thicker film for the determination of stoichiometry.
13

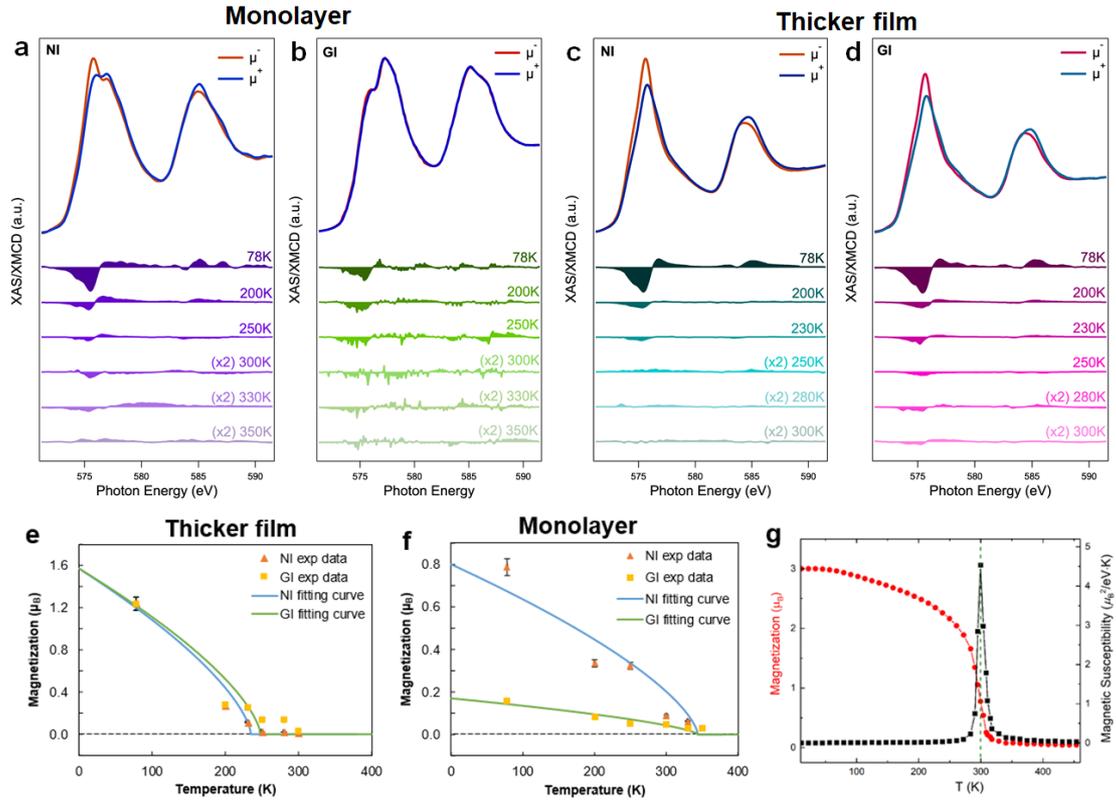

**Figure 4** Magnetic properties of $Cr_3Te_4$. (a-d) XAS and XMCD spectra of $Cr_3Te_4$ monolayer (1 u.c.) measured from 78 K to 350 K in (a) NI and (b) GI direction, and $Cr_3Te_4$ thick film (10 u.c.) measured from 78 K to 300 K in (c) NI and (d) GI direction. (e-f) The total magnetic moments of (e) $Cr_3Te_4$ thick film (10 u.c.) and (f) monolayer (1 u.c.) as a function of the temperature, where the experimental data were extrated from the XMCD specta, and the fiting M-T curves exhibit Curie-like behaviour. (g) Theoretical M-T curve of monolayer $Cr_3Te_4$ produced by Monte Carlo simulations, showing a room temperature $T_C$.

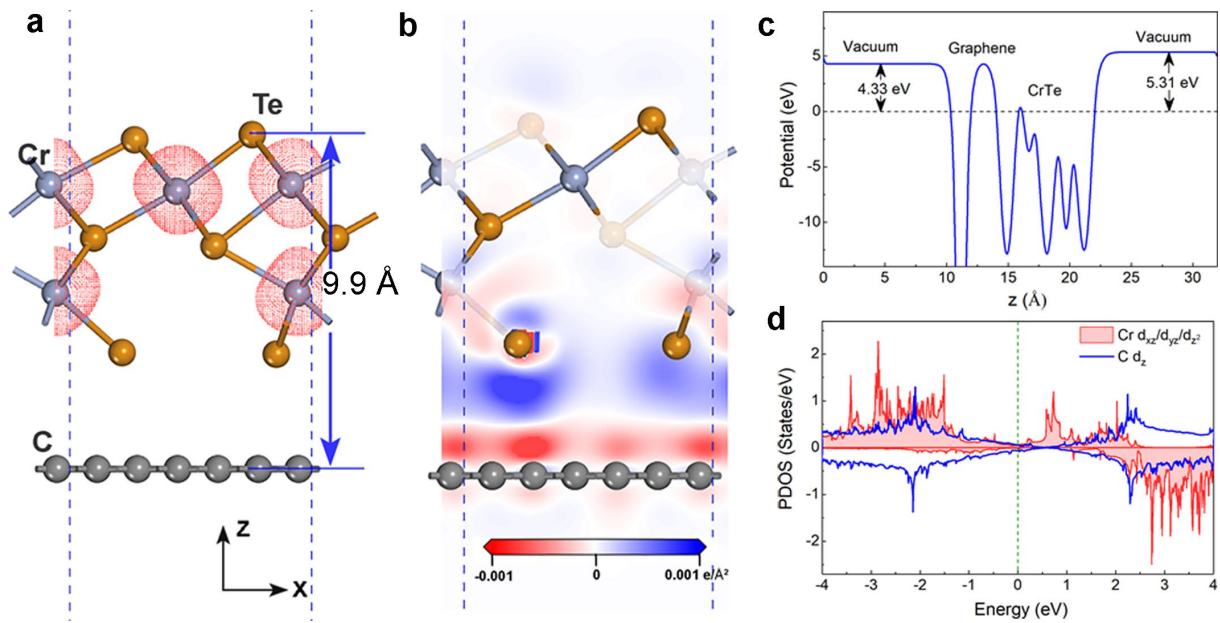

**Figure 5** DFT calculations of 2D $Cr_3Te_4$/graphene heterointerface. (a) Visualized spin-density with an isosurface value of $2×1.0^{-3}$ e/Å$^3$ for $Cr_3Te_4$ monolayer on graphene, where the red dots denote the spin density. (b) The contour plot of the charge density difference between the $Cr_3Te_4$ monolayer and graphene, in which the red and blue colour denote depleted and accumulated charge density, respectively. (c) In-plane averaged electrostatic potential across the interface of $Cr_3Te_4$/graphene, which is aligned to Fermi level. (d) The project density of states (PDOSs) on $d_{xz}$, $d_{yz}$, and $d_{z2}$ orbital of Cr and $p_z$ orbital of graphene for the Cr and C atoms.